\documentclass[svgnames,usenames,preprint,nocopyrightspace]{sigplanconf}

\usepackage{amsmath}

\usepackage{tikz}
\usetikzlibrary{arrows}
\usetikzlibrary{matrix}
\usepackage{listings}
\usepackage{amssymb}
\usepackage{xspace}
\usepackage[colorlinks=true]{hyperref}

\usepackage{xcolor}

\definecolor{dkgreen}{rgb}{0,0.6,0}
\definecolor{gray}{rgb}{0.5,0.5,0.5}
\definecolor{mauve}{rgb}{0.58,0,0.82}
\definecolor{light-red}{rgb}{1.0,0.8,0.8}
\definecolor{yellow-orange}{rgb}{1.0,0.85,0.05}
\definecolor{gray-blue}{rgb}{0.57,0.72,0.84}
\definecolor{light-gray}{gray}{0.80}

\definecolor{DarkGreen}{rgb}{0.0, 0.2, 0.13}

\definecolor{apricot}{rgb}{0.98, 0.81, 0.69}
\definecolor{bananayellow}{rgb}{1.0, 0.88, 0.21}
\definecolor{babyblueeyes}{rgb}{0.63, 0.79, 0.95}

\usepackage{soul}
\sethlcolor{titlebgcolor}

\newcommand\red[1]{\setlength\fboxsep{0pt}\colorbox{light-red}{\strut #1}}

\newcommand\grey[1]{\setlength\fboxsep{0pt}\colorbox{babyblueeyes}{\strut #1}}

\newcommand\RL{reinforcement\ learning\ }
\newcommand\ML{machine\ learning\ }

\lstdefinestyle{cstyle}{frame=b,
  escapechar={@},
  language=c,
  aboveskip=3mm,
  belowskip=3mm,
  showstringspaces=false,
  columns=flexible,
  basicstyle={\footnotesize\ttfamily},
  numbers=none,
  numberstyle=\tiny\color{gray},
  keywordstyle=\color{blue},
  keywordstyle=[2]\color{dkgreen},
  keywordstyle=[3]\color{magenta},
  commentstyle=\color{gray},
  stringstyle=\color{mauve},
  breaklines=true,
  breakatwhitespace=true,
  tabsize=3,
}

\lstdefinestyle{cstyleInline}{%frame=b,
  escapechar={@},
  language=c,
  aboveskip=3mm,
  belowskip=3mm,
  showstringspaces=false,
  columns=flexible,
  basicstyle={\footnotesize\ttfamily},
  numbers=none,
  numberstyle=\tiny\color{gray},
  keywordstyle=\color{blue},
  keywordstyle=[2]\color{dkgreen},
  keywordstyle=[3]\color{magenta},
  commentstyle=\color{gray},
  stringstyle=\color{mauve},
  breaklines=true,
  breakatwhitespace=true,
  tabsize=3,
  morekeywords={\#pragma}
}

\usepackage[textwidth=0.3\textwidth,textsize=scriptsize,colorinlistoftodos]{todonotes}

\begin{document}

\setlength{\pdfpageheight}{\paperheight}
\setlength{\pdfpagewidth}{\paperwidth}

\conferenceinfo{CONF 'yy}{Month d--d, 20yy, City, ST, Country}
\copyrightyear{20yy}
\copyrightdata{978-1-nnnn-nnnn-n/yy/mm}
\copyrightdoi{nnnnnnn.nnnnnnn}

% Uncomment the publication rights you want to use.
%\publicationrights{transferred}
%\publicationrights{licensed}     % this is the default
%\publicationrights{author-pays}

%\titlebanner{banner above paper title}        % These are ignored unless
%\preprintfooter{short description of paper}   % 'preprint' option specified.
\preprintfooter{PROHA'16, March 12, 2016, Barcelona, Spain}

\title{Towards Automatic Learning of Heuristics
for Mechanical Transformations of Procedural Code}
%\subtitle{Subtitle Text, if any}

\authorinfo{Guillermo Vigueras \and Manuel Carro}
           {IMDEA Software Institute \\ Campus de Montegancedo 28223 \\ Pozuelo de Alarc\'on, Madrid, Spain}
           {\{guillermo.vigueras,manuel.carro\}@imdea.org }
\authorinfo{Salvador Tamarit \and Julio Mari\~no}
           {Universidad Polit\'ecnica de Madrid \\ Campus de Montegancedo 28660 \\ Boadilla del Monte, Madrid, Spain}
           {\{salvador.tamarit,julio.marino\}@upm.es}

\maketitle

\begin{abstract}
%\gvcommin{Rewrite abstract}
%In this paper we describe an approach for learning heuristics to guide program transformations for Heterogeneous Systems. The approach uses different Machine Learning techniques like Classification and Reinforcement Learning. We show some preliminary results for some use case applications.
The current trend in next-generation exascale systems goes towards integrating a wide range of specialized (co-)processors into traditional supercomputers. However, the integration of different specialized devices increases the degree of heterogeneity and the complexity in programming such type of systems. %This high degree of heterogeneity makes programmability of such systems only possible for a few experts, and almost prohibits portability of the application onto different computational infrastructures.
Due to the efficiency of heterogeneous systems in terms of Watt and FLOPS per surface unit, opening the access of heterogeneous platforms to a wider range of users is an important problem to be tackled. In order to bridge the gap between heterogeneous systems and %non-expert 
programmers, in this paper we propose a \ML-based approach to learn heuristics for defining transformation strategies of a
program transformation system. Our approach proposes a novel combination of \RL and Classification methods to efficiently tackle the problems inherent to this type of systems. Preliminary results demonstrate the suitability of the approach for easing the programmability of heterogeneous systems.

\end{abstract}

% general terms are not compulsory anymore,
% you may leave them out


\section{Introduction}

%\gvcommin{State that the principles of the approach can be applied to other programming languages}
%\gvcommin{State that although the results are obtained for OpenCL, the approach can be used for other target platforms. Well do in the future}

The introduction of multi-core processors marked the end to an era
dominated by a constant increase of clock-rate, following Moore's law. Today, even the strategy to increase performance through higher core-counts is facing physical obstacles and, more importantly, power consumption issues which are currently one of the main limitations for the transition from petascale systems to next-generation exascale systems. Hence, the metric for measuring performance improvement of computational platforms is no longer absolute FLOPS numbers but rather FLOPS per Watt and FLOPS per surface unit of hardware. %As an example of these limitations, supercomputers performance has doubled more than 3000 times in the past 15 to 20 years, whereas the performance per watt has increased 300 times and performance per square foot has only increased 65 times in the same period of time~\cite{Younge:2010}. 

Thus, rather than having CPU-based supercomputers a growing trend goes towards integrating a wide range of specialized (co-)processors into traditional supercomputers. These specialized architectures can outperform general purpose ones while requiring lower energy consumption values and also less hardware real state. A study of energy consumption in data centers~\cite{Koomey2008}, has revealed the impact that efficient computer platforms can have in the worldwide energy consumption. This study analyzed historical energy consumption data for the time period 2000-2005 and 2005-2010. The study results revealed that in 2005 data centers used 0.97\% of the world’s total electricity consumption and in 2010 the value increased up to 1.3\%. However, the predicted value for 2010 based on the trends in the time period 2000-2005 would have been 2.2\% instead of 1.3\%, as reported. The study explained that the reduction was mainly a consequence of an increase in the use of energy-efficient computational platforms, showing the economical impact of using \textit{greener} architectures.

 However, the integration of different specialized devices increases the degree of heterogeneity of the system.
% to perform well for typical recurring tasks, rather than for all purposes
%. This makes perfect sense in embedded computing, where the processors are typically only applied for a specific application domain, yet with increasing flexibility of the domain, specialization may come as a hindrance, rather than a benefit. Vice versa, for the processor manufacturers, it gets increasingly difficult to identify which capabilities are sensibly supported by the processor. One implicit trend goes towards reconfigurability of the processors to allow for dynamic adaptation to the domain- specific requirements – either only for design purposes or actually within productive systems.
This high degree of heterogeneity
% and the implicit very strong deviation even on ISA level
 makes programmability of such systems only possible for a few
 experts, and almost prohibits portability of the application onto
 different resource infrastructures. Classical programming models are
 all designed for (a) Turing-inspired sequential execution and (b) von
 Neumann-like memory architecture, whereas a compiler typically
 optimizes code only for one (homogeneous) destination infrastructure
 at a time. Heterogeneous destinations typically require all programming
 tasks for hybrid architectures to be manually done  by the developer. In fact, no common current programming model manages to ease the programmability of heterogenous platforms and the portability of applications to this type of platforms while keeping good performance.
 
% parallelisation, distribution and adaptation and still remain performant.

It is crucial to note in this context, that programming models are only
likely to be taken up by the programming communities if they  (a) are coherent with classical approaches, (b) are easy and intuitive and (c) show clear performance / usability improvements. Accordingly, development of new programming models that hide low-level details and ease the programmability of the range of architectures available in heterogenous systems is slow, even if some programming models have already been proposed in the past for each individual architecture (e.g. OpenCL \cite{OpenCL2010}).

Opening the access of heterogeneous platforms to a wider spectrum of users is an important issue to be tackled. In past decades, different scientific areas have shown an enormous advance and development thanks to the use of massive parallel architectures to implement \textit{in-silico} models. The use of ``virtual labs" through computational platforms has allowed to better understand the physical phenomena under study, investigate a much wider range of solutions and drastically reduce the cost with respect to performing real experiments. Since the exascale computing era is pushed and driven by the computational requirements of scientific and industrial applications, new programming models, tools and compilers need to be released in order to open the access of exascale systems to a wider programming community and allow researchers from different communities to apply their findings to the society and improve human welfare.

In order to ease the programmability of heterogeneous platforms, we propose in this paper a \ML-based approach to learn heuristics to guide the code transformation process of a rule-based program
transformation system \cite{tamarit15:padl-haskell_transformation}. This type of systems pose different problems like the search-space exploration problem arising from the application of transformation rules in arbitrary orders or the definition of a stop criteria for the transformation system. For the latter we propose the use of classification trees and for the former we propose a novel approach based on \RL. We have performed a preliminary evaluation of the approach obtaining promising results that demonstrate the suitability of the approach for this type of transformation systems.

%the
%approach described in this document proposes the integration of effective transformation schemes into a rule-based program
%transformation toolchain \cite{tamarit15:padl-haskell_transformation}
%which accepts a program developed in C, commonly used in the
%computer science community, and with the help of user-provided annotations
%transforms the input program to generate code for
%heterogeneous platforms.

The rest of the paper is organized as follows. Section~\ref{sec:SoA} reviews previous work in the field of program transformation systems in general and previous approaches using \ML techniques. Section~\ref{sec:toolchain} describes the toolchain where our \ML-based approach is integrated. Section~\ref{sec:abstraction} describes the abstraction of code defined in order to apply \ML methods. Section~\ref{sec:Learning} describes the methods used to learn program transformation strategies. Later, Section~\ref{sec:Results} presents some preliminary results and finally Section~\ref{sec:conclusions} summarizes the conclusions and proposes future work.

\section{State of the Art}
\label{sec:SoA}

Rule-based program transformation systems support the formulation of basic code transformation steps as generic rules and arrange their automatic application. This scheme offers the flexibility of splitting complex code transformations in small steps, admitting efficient implementations that scale to large programs. By adding more rules, the transformation system can also increase its capabilities in a scalable way. Rule-based systems also allow to decouple the definition of transformation steps (i.e. rule applications) from the strategies followed to apply the rules. This decoupling provides more flexibility to try different transformation strategies and select the best one according to the purpose of the system \cite{Bagge03CodeBoost,Schupp2002}. Rule-based transformation has been used before to generate code for different computational platforms.

%% Paralelismo y FPGAs
The transformation of C-like programs an its compilation into synchronous architectures, like FPGAs, and asynchronous platforms, like multi-core CPUs, have been addressed before~\cite{Brown2005-tr-opt_trans_hw}. However, the input language of the approach (Handel-C) is meant to specify synchronous systems, thus limiting its applicability to this type of systems.
%% Paralelismo y álgebra lineal
A completely different approach is to use linear algebra to transform the
mathematical specification of concrete scientific
algorithms~\cite{Franchetti2006,Fabregat2013,DiNapoli2014}.
Here, the starting point is a mathematical formula and, once the formula is
transformed, code is generated for the resulting expression. 
%While these works have shown
However, a reasonable acceleration over hand-tuned code 
happens %is expected 
only for those algorithms, and applying the ideas to
other contexts does not seem straightforward.

Machine learning techniques have been already used in the field of compilation and program transformation \cite{Mariani:2014,Pekhimenko:2010,Agakov:2006}. All these approaches share the same principles as they obtain an abstract representation of the input programs in order to apply \ML methods. Nevertheless, previous approaches target some specific architecture limiting the applicability of the approach and making it not suitable for heterogeneous platforms. In our approach, we obtain program abstractions but in order to enable the \ML-based transformation and compilation of a program written in C for heterogeneous systems. Additionally, none of the previous works have explored the use of \RL %(RL) \stcommin{The acronym is not used.} 
methods \cite{RLSurvey1996} in the field of program transformation and compilation.

\section{Program Transformation Toolchain for Heterogeneous systems}
\label{sec:toolchain}

In this work we propose the automatic learning of transformation strategies for a rule-based transformation toolchain \cite{tamarit15:padl-haskell_transformation,tamarit16:code_trans}. This type of systems offer a higher flexibility over compilation infrastructures where the code transformations and optimizations are embedded in the compilation engine \cite{Stallman2009}. Rule-based systems support the formulation of basic code transformation steps as generic rules and arrange their automatic application. This scheme offers the flexibility of splitting complex code transformations in small steps, admitting efficient implementations that scale to large programs. By adding more rules, the transformation system can also increase its capabilities in a scalable way. Rule-based systems also allow to decouple the definition of transformation steps (i.e. rules) from the strategies followed to apply the rules. This decoupling provides more flexibility to try different transformation strategies and select the best one according to the purpose of the system \cite{Bagge03CodeBoost,Schupp2002}.

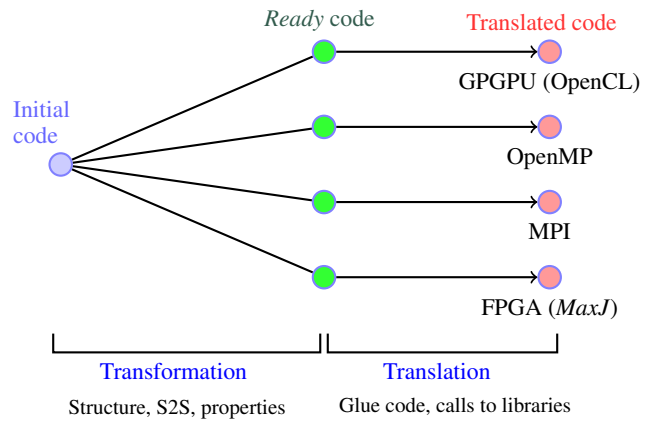
\begin{figure}[h!]
  \begin{center}
\begin{tikzpicture}[
    place1/.style={circle,draw=blue!50,fill=blue!20,thick},
    place2/.style={circle,draw=blue!50,fill=green!80,thick},
    place3/.style={circle,draw=blue!50,fill=red!40,thick},
    thick]
    \node (original) at (-1.5,0) [place1,label=above:\parbox{4em}{\textcolor{blue!60}{Initial code}}] {};
    \node (rgpu) at (2,1.5)
    [place2,label=above:\parbox{5em}{\textcolor{DarkGreen!80}{\emph{Ready} code}}] {};
    \node (romp) at (2,0.5) [place2] {};
    \node (rmpi) at (2,-0.5) [place2] {};
    \node (rfpga) at (2,-1.5) [place2] {};
    \node (gpu) at (5,1.5) [place3,label=below:GPGPU (OpenCL),label=above:\parbox{7em}{\textcolor{red!80}{Translated code}}] {};
    \node (omp) at (5,0.5) [place3,label=below:OpenMP] {};
    \node (mpi) at (5,-0.5) [place3,label=below:MPI] {};
    \node (fpga) at (5,-1.5) [place3,label=below:FPGA (\emph{MaxJ})] {};
    \draw [->] (original) -- (rgpu) -- (gpu);
    \draw [->] (original) -- (romp) -- (omp);
    \draw [->] (original) -- (rmpi) -- (mpi);
    \draw [->] (original) -- (rfpga) -- (fpga);
    \draw (-1.6,-2.25) -- (-1.6,-2.50) -- (1.95, -2.50) -- (1.95, -2.25);
    \draw (2.05,-2.25) -- (2.05,-2.50) -- (5.1, -2.50) -- (5.1, -2.25);
    \node [blue] at (0,-2.75) {Transformation};
    \node [blue] at (3.5,-2.75) {Translation};
    \node at (0.5,-3.25) {\parbox{12em}{\footnotesize\flushleft Structure, S2S, properties}};
    \node at (4.1,-3.1) {\parbox{12em}{\footnotesize\flushleft Glue code, calls to libraries}};
  \end{tikzpicture}
  \end{center}
  \caption{Stages of the transformation tool}
  \label{fig:ana-trans-tool}
 \end{figure}

The rule-based transformation toolchain mentioned before \cite{tamarit15:padl-haskell_transformation,tamarit16:code_trans} defines a
transformation process consisting of two main stages, as shown in
Figure~\ref{fig:ana-trans-tool}. The first one is a transformation
phase in which the input C code is transformed into 
semantically equivalent c Code but which is better suited for a given
platform. It 
basically reshapes the input code taking into account syntactic/semantic
restrictions of the compilers/programming models of each target
platform (e.g.\ if the target compiler does not accept general
\texttt{\textbf{if}} statements, the input code will be transformed to remove them if possible, without modifying the program semantics). %The programming models currently supported are: OpenMP and MPI for shared and distributed memory multi-cores, OpenCL for GPUs and MaxJ\footnote{https://www.maxeler.com/products/software/maxcompiler} for FPGAs.
In the first stage, transformation rules are formalized in an internal domain-specific language that takes care of the syntactic and semantic conditions required to apply a given transformation.  Transformation rules operate at \emph{abstract syntax tree} (AST) level transforming an input code that matches a pattern into another code, if rule conditions are met. In this way, this first phase defines a search space exploration problem where the different nodes of the search space would be the different states of the input code obtained as a result of applying code transformations and the transitions among the search space nodes would be the transformation rules applied. Thus, the high number of rules in the transformation system results in a combinatorial explosion of the code states to explore.
 
% Once the transformation stage is finished the second phase starts.
The second phase of the toolchain in Figure~\ref{fig:ana-trans-tool} consists of a translation process which generates code for a given platform taking as input the code generated in the first phase. Once platform-specific code is generated, it is compiled using a platform-specific compiler. This second phase poses another problem to be tackled: the definition of the starting condition for the translation phase, which is also the stop condition of the transformation process. Since depending on program complexity some platforms might require a long compilation process taking up to tens of hours, like for an FPGA, the toolchain should translate and compile only code that is likely to be successfully compiled on a given target platform rather than trying to compile every single code obtained during the transformation phase. Additionally, when considering the first and second phases together, the transformation sequences obtained in the first phase from the chaining of several rules, might not improve monotonically non-functional metrics associated to an input code (e.g. execution time, energy consumption, etc.), but still produce code with better performance in the second phase.

Given the nature of the problems associated with the transformation toolchain, we propose the use of different \ML techniques to effectively guide the transformation process. On one hand, we have used \RL methods to learn transformation strategies that take into account the non-monotonic behavior of the transformation sequences. On the other hand, we have used classification methods to learn appropriate states that can be translated for the target platforms. Also \ML techniques require descriptions of the problem domain. For that reason, we compute an abstract representation of the input programs and the intermediate codes generated during the transformation process.

 %decide which codes from the first phase are suitable to be translated in the second phase and then compile the translated code for the target platform.

%  But this second phase cannot be implemented as a ``trial and error" process in order to discard the platform specific codes not accepted by the platform compiler. Since some platforms require a compilation process that might take from some tens of minutes up to some tens of hours, like a FPGA, only those platform specific codes that can be successfully compiled should be generated.

%\begin{itemize}
%
%%\item Talk about the transformation and translation phases of the toolchain
%
%\item Describe the problems associated with the transformation strategies:
%\begin{itemize}
%\item Rule-based transformation $\Longrightarrow$ state-space exploration problem.
%\item Non-monotonic transformation sequences
%\item Determine final states
%\end{itemize}
%
%\end{itemize}

\section{Code-Abstraction Mapping}
\label{sec:abstraction}

As mentioned in previous section, \ML methods operate on descriptions of the problem domain. In our case, we need to obtain abstractions of procedural code written in C. These abstractions should reflect the code changes performed by the transformation rules which operate at the AST level. At the same time, mapping of code to abstractions must be univocal in order to avoid conflicts and ensure a correct behavior of \ML methods. For that reason, we represent abstractions as quantitative descriptions that capture changes performed by rules on code. These quantitative descriptions involve features like: AST patterns, control flow, data layout, data dependencies, etc. To obtain these abstractions we have developed a 
static code analysis (SCA) 
tool that parses the AST to extract the features mentioned before. The SCA tool has been implemented in Python using the \textit{pycparser}\footnote{\url{https://github.com/eliben/pycparser}} module and can obtain feature information either by directly analysing the code or by parsing code annotations provided by the user or by third-party analysers. The set of code annotations considered in the abstractions coincides with the annotations supported by the transformation toolchain mentioned before~\cite{tamarit15:padl-haskell_transformation}.

Currently, code abstractions consist of a vector formed out of fifteen features. Some of them capture changes in code performed by transformation rules and also reflect code patterns that match the syntactic/semantic restrictions of target compiler/programming models:

%\gvcommin{describe features}
%\gvcommin{justify why we have identified these features so far $\rightarrow$ Reason: because is sufficient for the use cases we have considered. It will be extended in the future}

\begin{itemize}
\item{\textbf{Maximum nested loop depth:}} This feature measures the depth of nested \texttt{for} loops. A value of $0$ for this feature means no nested loops are present, a value of $1$ means two nested loops, a value of $2$ means three nested loops and so on.

\item{\textbf{Number of function calls:}} This feature counts the number of function calls present in the analysed code.

\item{\textbf{Number of array writes shifted within for loops:}} This feature counts the number of array accesses with positive offset for the following pattern: a \texttt{for} loop with writes accesses of multiple positions of the same array within the same loop iteration and some of the write accesses use a positive offset with respect to the loop iteration variable.

The following two codes explain the meaning of this feature. For the code on the left, this feature would take a value of 1 since for each iteration there are multiple write accesses, one of them with positive offset (i.e. \texttt{i+1}). However, for the code on the right this feature takes a value 0 since for each iteration only position \texttt{i+1} is written even if it is done with positive offset. The multiple writes within the same loop iteration is detected by this feature since a code with this pattern will fail to be compiled for a FPGA. %using the MaxJ programming model that we currently use to target FPGAs.

% \begin{center}
\begin{minipage}{0.23\textwidth}
\begin{lstlisting}[style=cstyleInline,xleftmargin=.09\textwidth]
for(i=1;i<N;i+=2) {
   v[i] = v[i-1];
   v[i+1] = v[i-1]*i;
}
\end{lstlisting}
\end{minipage}
\begin{minipage}{0.23\textwidth}{
\begin{lstlisting}[style=cstyleInline,xleftmargin=.09\textwidth]
for(i=0;i<N-1;i++) {
   aux = i*i;
   v[i+1] = aux;
}     
\end{lstlisting}
}
\end{minipage}     
% \end{center}

\item{\textbf{Irregular loops:}} This feature takes a binary value and informs whether loops contain ($1$) or not ($0$) statements like \texttt{break} or \texttt{continue} that alter the normal flow a program.

\item{\textbf{Global variables:}} This feature takes a binary value and states if any global variable is written within the piece of code analysed ($1$) or not ($0$).

\item{\textbf{If statements:}} This feature counts the number of \texttt{if} statements within the piece of code analysed.

\item{\textbf{Static limits of for loop:}} This feature informs if all the \texttt{for} loops in the piece of code analysed do not have static iteration limits, i.e. if they do not change along the iteration space of the loop. The following code shows an example where the upper iteration limit of the inner loop is not static since it might change for each iteration of the outer loop if elements of array \texttt{v} are removed by function \texttt{clean}. 

% \begin{center}
\begin{lstlisting}[style=cstyleInline,xleftmargin=.13\textwidth]
for(j=0;j<N;j++) {     
   for(i=0;i<size(v);i++)
      update(v[i]);
           
   clean(v);
}
\end{lstlisting}
% \end{center}

\item{\textbf{Iteration independent loop:}} This feature counts the number of \texttt{for} loops in the code analysed without carried dependencies across iterations. This information is obtained from annotations provided by the user or by third-party code analysers.

\item{\textbf{Any for loop with loop\_schedule:}} This feature detects if there is some pattern composed of two nested loops used to iterate over an array split in chunks. This code pattern is detected through an annotation that can be inserted either by the user or by the transformation tool after applying a loop scheduling transformation. The following code shows an example of loop\_schedule where array \texttt{v} is accessed in chunks of size N: %Loop scheduling divides a loop into multiple parts that may be run concurrently on multiple processors.

\begin{lstlisting}[style=cstyleInline,xleftmargin=.13\textwidth]
#pragma stml loop_schedule
for(j=0;j<M;j++) {     
   w[j] = 0;
   for(i=0;i<N;i++)
      w[j] += v[j*N+i];
}
\end{lstlisting}

\item{\textbf{Number of loop invariant var:}} This feature quantifies the number of variables that are assigned outside a \texttt{for} loop and are not modified within it.

\item{\textbf{Number of loop hoisted var modifications:}} This feature is used to count the number of variables that are assigned outside a \texttt{for} loop and are modified within the loop.

\item{\textbf{Number of non-1D array:}} This feature counts the number of arrays in the code with a number of dimensions higher than 1.

\item{\textbf{Number of auxiliary variables to access arrays:}} This feature counts the number of auxiliary variables used to index an array. If a variable used as array index is not within the set of iteration variables of the \texttt{for} loops that iterates over the array, then it is considered as an auxiliary index variable. The following code shows a simple example where the \texttt{aux} variable is used to index the array \texttt{v} instead of using the iteration variable \texttt{j} of the \texttt{for} loop:

\begin{lstlisting}[style=cstyleInline,xleftmargin=.13\textwidth]
aux = 0;
for(j=0;j<N;j++) {     
   w[i] = v[aux];
   aux++;
}
\end{lstlisting}

\item{\textbf{Total number of for loops:}} This feature counts the total number of \texttt{for} loops independently of the nesting of loops.

\item{\textbf{Non-normalized for loops:}} This feature counts the number of non-normalized \texttt{for} loops. Currently, a \texttt{for} loop is considered as normalized if the step of the iteration is $1$ and it is considered as non-normalized otherwise.

%\begin{itemize}
%\item ML generally operates on \emph{descriptions} of real world.
%\item Applying ML to program transformation through abstractions of programs.
%\begin{itemize}
%\item Based on code features related with:
%\begin{itemize}
%\item Control flow,
%\item Data layout,
%\item Data dependencies\ldots
%\item Also on code annotations (externally provided.)
%
%\end{itemize}
%
%\item Abstractions should avoid \emph{state conflicts}: substantially
%  different code fragments are mapped to different abstractions.
%\item Note: previous works apply ML to e.g.\ compilation.
%\end{itemize}
%\end{itemize}
%
\end{itemize}
 %Regardless the code abstractions obtained by
The SCA tool  
%implemented in Python 
in charge of obtaining the abstractions described above can be seen as a function mapping codes to abstractions, defined in the following way:

\[A:\ Code \rightarrow Abstraction\]

Function $A$ is used later in Section~\ref{sec:RL} for explaining the mapping of codes to \RL states. Currently code abstractions consist of the fifteen features described before since they were sufficient to obtain some preliminary results for the set of use case applications described in Section~\ref{sec:Results}. However, we plan to increase the vector of features as we increase the set of use case applications.

\section{Automatic Learning of Transformation Heuristics}
\label{sec:Learning}

%\begin{itemize}
%\item Decision trees to classify final states
%\item Reinforcement Learning to find transformation sequences
%
%\item Simple example (from slides)
%\end{itemize}

As mentioned before, the transformation engine that guides the toolchain must tackle some problems associated with rule-based code transformation systems. On one hand, the search space with non-monotonic behavior must be efficiently explored and on the other hand an effective stop criteria for the search procedure must be defined. Our approach uses \RL to solve the former problem and classification trees to solve the latter. Following subsections describe the details of these \ML techniques. An example to show how it works in practice is provided later.

\subsection{Classification Trees}
\label{sec:CT}

In \ML and statistics, classification is the problem of identifying the category to which a new observation belongs among a set of pre-defined categories. The classification is done on the basis of a training set of data containing observations for which it is known to which category they belong~\cite{Marsland2009}. Different formalisms can be used to classify. We have decided to start evaluating the adequacy of classification trees for our problem since it intuitively and implicitly performs feature selection without hard data preparation requirements.

A classification tree is a simple representation for classifying examples according to a set of input features. All of the input features have finite discrete domains, and there is a single target variable called the \textit{classification} feature. Each element of the domain of the target variable is called a class. In a classification tree each internal (non-leaf) node is labeled with an input feature. Each leaf of the tree is labeled with a class or a probability distribution over the classes. Thus, a tree can be learned by splitting the source data set into subsets based on values of input features. This process is repeated on each derived subset in a recursive manner called recursive partitioning. The recursion is completed when the subset at a node has the same value of the target variable, or when splitting no longer improves the predictions. Typically, the source data comes in records of the form: %This process of top-down induction of decision trees (TDIDT) is an example of a greedy algorithm, and it is by far the most common strategy for learning decision trees from data.

\[(\textbf{x},Y) = ([x_1, x_2, x_3, ..., x_k], Y)\]

The dependent variable, $Y$, is the target variable that the classification tree generalizes in order to be able to classify new observations. The vector $\textbf{x}$ is composed of the input features $x_i$, used for the classification. In this context, the source input data for our problem is composed of the vectors of the abstractions described in Section~\ref{sec:abstraction}, i.e., $k=15$ in our case. The domain of the target variable can take values among the four different final platforms we currently support: FPGA, GPU, Shared-Memory CPU (SM-CPU) and Distributed-Memory CPU (DM-CPU). Since a given code and its associated abstraction might be well suited for more than one platform, we obtain 15 possible classes for our target variable. This number is computed as the sum all the combinations of four elements taken $m$ at a time, where $1 \le m \le 4$. So the total number of classes $N$ is computed as:

\[ N = \sum_{m=1}^{4}C_{4}^{m} = 15,\ \emph{where}\ C_{n}^{m}=\frac{n!}{m!(n-m)!} \]

The classes obtained for the target variable allow to define the final states of the transformation stage of the toolchain described in Section~\ref{sec:toolchain}, which will also serve to define the final states for the \RL algorithm that is described in the following section. As a final remark, the classification-based learning described in this section have been implemented using the Python library \textit{Scikit-learn} \cite{scikit2011}. This library implements several \ML algorithms and is widely adopted in the scientific community providing good support and ample documentation.

\subsection{Reinforcement Learning}
\label{sec:RL}

Reinforcement learning~\cite{Marsland2009} is an area of \ML
%inspired by behaviorist psychology, 
%\mclcomm{I wouldn't mention psychology / behavior...}
 concerned with how software agents ought to take actions in an environment so as to maximize some notion of cumulative reward.  A \RL agent interacts with its environment in discrete time steps. At each time $t$, the agent receives an observation $o_t$, which typically includes the reward $r_t$. It then chooses an action $a_t$ from the set of available actions, which is subsequently sent to the environment. The environment moves from current state $s_t$ to a new state $s_{t+1}$ providing the reward $r_{t+1}$ associated with the transition $(s_t,a_t,s_{t+1})$. The goal of a \RL agent is to collect as much reward as possible. 

According to the previous description, \RL seems well suited to represent the optimization process of a programmer or a compiler, since it typically consists of
iteratively improving  an initial program in discrete steps, where code changes correspond to actions and code versions obtained during the optimization process correspond to states.  Moreover, code is typically evaluated after every change, often  according to some non-functional properties such as execution time, memory consumption speedup factor, \ldots  The result of these evaluations
%% is used to the concept of delayed reward where the measurements of
%% these non-functional properties
can be easily translated into rewards and penalties that support the
learning process.

The result of the learning process of the agent is a \textit{state-action} table (Figure~\ref{fig:sa-table}) which will eventually contain values for each combination $(s, a)$ of states and actions.   These values are scores for the expected profit to be obtained from applying action $a$ to state $s$.  This table is initially filled in with a default value and is iteratively updated following a learning process which we briefly describe below.

%% is initialized to some default value and wh. This 2D table
%% (Figure~\ref{fig:sa-table}), containing as many rows as states and
%% as many columns as actions

The process of \RL is based on a set of predetermined transformation sequences, which are assumed to be models to learn from.  Each sequence $S$ is composed of a set of states $S = s_0, s_1, \ldots, s_{t-1}, s_t, s_{t+1}, \ldots, s_l$ and the actions which transform one state into the next one.
% and the associate rewards, provided as a training set.  
The final state of each transformation sequence has a different reward value related, in our case, with the performance of the final code corresponding to state $s_l$ in the sequence (better performance gives higher rewards).  The \RL training phase consists of an iterative, stochastic process in which states (i.e. code abstractions) from the training set are randomly selected. For each  state $s$ a \emph{learning episode} is started by selecting the action $a$ with the highest value in $Q$ for that $s$ and moving to a new state $s'$ according to transition $(s,a,s')$. From state $s'$ the same transition process is repeated to advance in the learning episode until a final state is reached or a given number of episode steps is performed. When the episode terminates, the values in $Q$  corresponding to the states and actions of the visited sequence are updated according to Equation~\ref{eq:RL}, where $Q_{init}(s_t,a_t)$ is the initial value of $Q$ for state $s_t$ and action $a_t$.   Note that $s_t$ (resp.\ $a_t$) is not the $t$-th state in some fixed ordering of states, but the $t$-th state in the temporal ordering of states in the sequence used to learn.

% , is updated according to the following recursive expression:
 
 \begin{figure}
   \centering

\begin{tikzpicture}[mycell/.style={minimum size=1cm}]
\draw (-2,-3) -- (3,-3);
\draw (-2,-2) -- (3,-2);
\draw (-2,0) -- (3,0);
\draw (-2,1) -- (3,1);
\draw (-2,2) -- (3,2);
\draw (-2,2) -- (-2,-3);
\draw (-1,2) -- (-1,-3);
\draw (0,2) -- (0,-3);
\draw (2,2) -- (2,-3);
\draw (3,2) -- (3,-3);
\matrix (A) [matrix of math nodes,row sep=-\pgflinewidth, column
sep=-\pgflinewidth,nodes={mycell}] {
    & a0      & a1 &        &        & an \\
\node (s0) {s0}; & q_{0,0}      & q_{0,1} & \cdots & \cdots & q_{0,n} \\
\node (s1) {s1}; & q_{1,0}      & q_{1,1} &        &        & q_{1,n} \\
    & \vdots &   &        &        & \vdots \\
    & \vdots &   &        &        & \vdots \\
\node (sm) {sm}; & q_{0,m}      & q_{1,m} & \cdots & \cdots & q_{m,n} \\
};
 \node (pr) [color=gray,left of=A-4-2,xshift=-10]
     {\rotatebox{180}{\resizebox{5mm}{4cm}{\}}}};
 \node [xshift=-100,yshift=-15] (left of=pr) {States};
 
 \node (pu) [color=gray,above of=A-2-3,xshift=30,yshift=15]
     {\rotatebox{90}{\resizebox{5mm}{4cm}{\}}}};
 \node  [xshift=15,yshift=100] (above of=pu) {Actions};

\end{tikzpicture}

\caption{State-Action table $Q$.  It will eventually be filled in with
  values $q_{i,j} \in \mathbb{R}$ obtained from the learning process.}
   \label{fig:sa-table}
 \end{figure}

 %The agent can choose any action as a function of the history and it can even randomize its action selection.

%\[Q(s_t,a_t) \leftarrow Q(s_t,a_t) + \alpha [r_{t+1} + \gamma Q(s_{t+1}, a_{t+1})-Q(s_t,a_t)]\]

%\small
\begin{equation}
Q(s_t,a_t) = \left\{
\begin{array}{ll}
Q(s_t,a_t) + \alpha~\cdot~(r_{t+1} & $if $ s_t\ not\ final \\
~~+ ~\gamma~\cdot~Q(s_{t+1}, a_{t+1})  & \\
~~- ~Q(s_t,a_t)~) & \\
                              & \\
             Q_{init}(s_t,a_t) & $otherwise$
\end{array}
\right.
%Q(s_t,a_t) = 
%  \begin{cases} 
%  Q(s_t,a_t) +  & s_t\ not\ final \\
%\alpha [r_{t+1} + \gamma Q(s_{t+1}, a_{t+1}) -Q(s_t,a_t)]  &  \\
%  & \\
%   Q(s_t,a_t)       & otherwise
%  \end{cases}
%  \label{eq:RL}
\label{eq:RL}                                                                             
\end{equation} 

%\stcommin{There are infinite loop in this equation, no?}

The final states in Equation~\ref{eq:RL} are defined based on the classification as described in Section~\ref{sec:CT}. Two additional parameters appear in Equation~\ref{eq:RL}: the learning rate $\alpha, 0 < \alpha \le 1$, and the discount factor $\gamma, 0 < \gamma \le 1$. The learning rate determines to what extent the newly acquired information will override the old information. A factor of 0 will make the agent not learn anything, while a factor of 1 would make the agent consider only the most recent information.  The discount factor implements the concept of \textit{delayed reward} by determining the importance of future rewards. A factor of 0 will make the agent opportunistic by considering only current rewards, while a factor close to 1 will make it strive for a long-term high reward.  If the discount factor reaches or exceeds 1, the values in $Q$  may diverge~\cite{Marsland2009}.
  
%has been used in different fields where the learning process is guided by rewards obtained from environmental observations resulting from actions taken to change the environment.

In order to use the \RL \textit{state-action} in our setting, 
%table once the training is finished, 
we need to define some mappings.  The abstraction of a concrete piece of code is provided by function $A$ (Section~\ref{sec:abstraction}).  Abstractions and transformation rules must be mapped to states and actions, respectively, in order to index the \textit{state-action} table. This mapping is done through functions $SM$ and $AM$, defined as:

\[SM:\ Abstraction \rightarrow State\]
\[AM:\ Rule \rightarrow Action\]

\medskip

%\begin{equation*}
%\begin{array}{ll}
%SM:  & Abstraction \rightarrow State \\
%        &     \\
%AM:  & Rule \rightarrow Action
%\end{array}
%\end{equation*}

Based on the mapping of abstractions and rules defined, the \RL \textit{state-action} table of Equation \ref{eq:RL} can also be modeled as a function $Q$:

\[Q:\ State\ \times\ Action \rightarrow \mathbb{R}\]

\medskip

%\begin{equation*}
%\begin{array}{ll}
%\ Q:  & State\ \times\ Action \rightarrow \mathbb{R}
%\end{array}
%\end{equation*}

Using functions $A$, $SM$, $AM$ and $Q$, the strategy of the transformation toolchain for selecting rules at each transformation step can be modeled with a function $RS$:

\[RS:\ Code\ \rightarrow Rule\]

\medskip

%\begin{equation*}
%\begin{array}{ll}
%RS:  & Code\ \rightarrow Rule
%\end{array}
%\end{equation*}

This function takes as input a given code $c$ and selects the transformation rule $ru$ associated to action $AM(ru)$, that maximizes the value provided by $Q$ for the state $SM(A(c))$ associated to input code $c$. Thus, the rule selection strategy can be expressed as:

\[RS(c)\ =\ \operatorname*{arg\,max}_{ru \in Rule} Q(SM(A(c)),AM(ru))\]

%\[\underset{x\in S}{\operatorname{arg\,max}}\, f(x) := \{x \mid x\in S \wedge \forall y \in S : f(y) \le f(x)\}\]

The operator $\operatorname*{arg\,max}$ may return, by definition, the empty set, a singleton, or a set containing multiple elements. However, in our problem parameters $\alpha$ and $\gamma$ as well as the reward values $r_{t+1}$
%\stcommin{Symbol $r$ is being used to represent rules and rewards. This can be confusing. In Section \ref{sec:example} symbol $R$ is used to represent rules, so maybe is better to use this symbol also here. }
 appearing in Equation~\ref{eq:RL} can be tuned to ensure that a single rule is returned, this avoiding a non-deterministic $RS$ function. Section~\ref{sec:Results} gives further details on how we selected their values. % selected for all these \RL parameters. 
%\gvcommin{Decir que en resultados explicamos los valores usados para alfa, gamma y reward}

 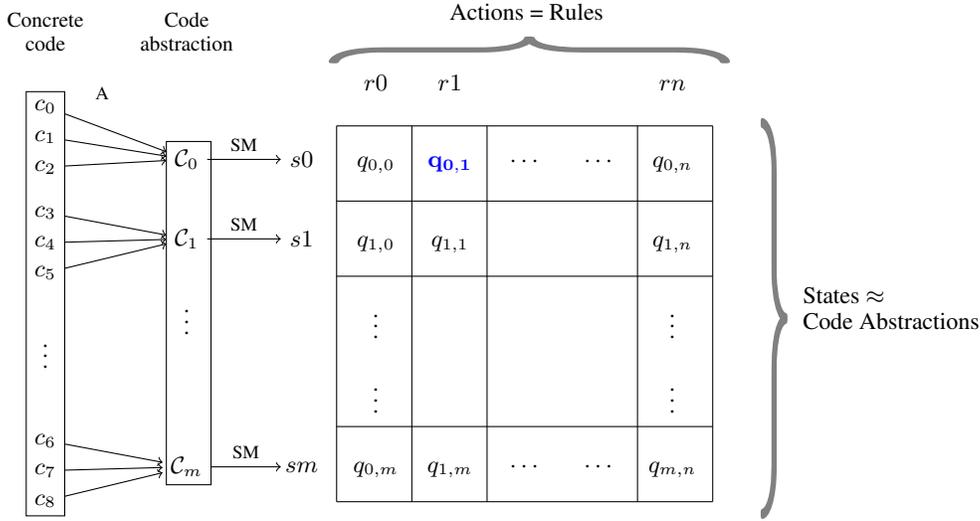
\begin{figure*}[]
   \centering
\begin{tikzpicture}[mycell/.style={minimum size=1cm}]

%\draw[shift={(-1.5,-2.5)}] (0,0) grid (4,4);
%\draw[shift={(-1.5,0.5)}] (0,0) grid (4,2);
\draw (-2,-3) -- (3,-3);
\draw (-2,-2) -- (3,-2);
\draw (-2,0) -- (3,0);
\draw (-2,1) -- (3,1);
\draw (-2,2) -- (3,2);
\draw (-2,2) -- (-2,-3);
\draw (-1,2) -- (-1,-3);
\draw (0,2) -- (0,-3);
\draw (2,2) -- (2,-3);
\draw (3,2) -- (3,-3);
\matrix (A) [matrix of math nodes,row sep=-\pgflinewidth, column
sep=-\pgflinewidth,nodes={mycell}] {
    & r0      & r1 &        &        & rn \\
\node (s0) {s0}; & q_{0,0}      & \textcolor{blue}{\ensuremath{\mathbf{q_{0,1}}}} & \cdots & \cdots & q_{0,n} \\
\node (s1) {s1}; & q_{1,0}      &  q_{1,1} &      &        & q_{1,n} \\
    & \vdots &   &        &        & \vdots \\
    & \vdots &   &        &        & \vdots \\
\node (sm) {sm}; & q_{0,m}      & q_{1,m} & \cdots & \cdots & q_{m,n} \\
};
\node (c0) [left of=s0,xshift=-15] {$\mathcal{C}_0$};
\node (c1) [left of=s1,xshift=-15] {$\mathcal{C}_1$};
\node (cm) [left of=sm,xshift=-15] {$\mathcal{C}_m$};
\draw [->] (c0) -- ++(1.25,0) node [midway,above] {\scriptsize SM};
\draw [->] (c1) -- ++(1.25,0) node [midway,above] {\scriptsize SM};
\draw [->] (cm) -- ++(1.25,0) node [midway,above] {\scriptsize SM};

\node (rc0) [left of=c0, yshift=20,xshift=-25] {$c_0$};
\node (rc1) [below of=rc0,yshift=17] {$c_1$};
\node (rc2) [below of=rc1,yshift=17] {$c_2$};
\draw [->] (rc0) -- (c0);
\draw [->] (rc1) -- (c0);
\draw [->] (rc2) -- (c0);
\node [right of=rc0,xshift=-7,yshift=5] {\scriptsize A};

\node (rc3) [left of=c1, yshift=10,xshift=-25] {$c_3$};
\node (rc4) [below of=rc3,yshift=17] {$c_4$};
\node (rc5) [below of=rc4,yshift=17] {$c_5$};
\draw [->] (rc3) -- (c1);
\draw [->] (rc4) -- (c1);
\draw [->] (rc5) -- (c1);

\node (rc6) [left of=cm, yshift=10,xshift=-25] {$c_6$};
\node (rc7) [below of=rc6,yshift=17] {$c_7$};
\node (rc8) [below of=rc7,yshift=17] {$c_8$};
\draw [->] (rc6) -- (cm);
\draw [->] (rc7) -- (cm);
\draw [->] (rc8) -- (cm);

\node [below of=rc5] {\vdots};
\node [below of=c1] {\vdots};

\draw (rc0.north west) rectangle (rc8.south east);
\node [above of=rc0] (ac) 
    {\parbox{4em}{\centering\footnotesize Concrete \\ code}};

\node [above of=c0, yshift=20]
    {\parbox{4em}{\centering\footnotesize Code \\ abstraction}};
\draw (c0.north west) rectangle (cm.south east);

%% \node [above of=s0, yshift=20]
%%     {\parbox{4em}{\centering\footnotesize States}};

\node (pr) [color=gray,right of=A-4-6,xshift=10]
    {\resizebox{5mm}{4cm}{\}}};
\node [xshift=160,yshift=-12] (right of=pr) {\parbox{9em}{States $\approx$ \\ Code Abstractions}};

\node (pu) [color=gray,above of=A-2-3,xshift=30,yshift=15]
    {\rotatebox{90}{\resizebox{5mm}{4cm}{\}}}};
\node  [xshift=15,yshift=100] (above of=pu) {Actions = Rules};

\end{tikzpicture}

   \caption{State-Action table for code, code abstraction, and rules.}
   \label{fig:sa-code-rules}
\end{figure*}

%When using the table synthesized through reinforcement learning, 
The relationships among the code, its abstraction, the rules, and the
contents of the state-action matrix are depicted in
Figure~\ref{fig:sa-code-rules}.  Table $Q$ is used as follows: for a
concrete code $c_k$ we find its abstraction $\mathcal{C}_i = A(c_k)$.
Let us assume $i=0$.  From the row $i$ corresponding to
$\mathcal{C}_i$ in matrix $Q$ we obtain the column $j$ with the
highest value $q_{i,j}$ (in our example, $q_{0,1}$, in blue and
boldface).  Column $j$ corresponds to rule $R_j$, which is expected to
give the most promising code when applied to a code state whose
abstraction is $\mathcal{C}_i$ (in our case it would be $R_1$).  Rule
$R_j$ would be applied to $c_k$ to give $c'$.  If $c'$ corresponds to
a final state, the procedure finishes.  Otherwise we repeat the
procedure taking $c'$ as input and finding again a rule to transform
$c'$.

% Since each transformation sequence is composed of steps for each of which a code is transformed by a rule application, 
% the transformation rules of each sequence will be learnt as the most promising actions of
%\gvcommin{Explicar las secuencias de transformacion, iteraciones y episodios de aprendizaje, sampling aleatorio}

 In order to implement the \RL-based module in charge of learning heuristics for program transformation we have used the Python library \textit{PyBrain}~\cite{pybrain2010}. This library adopts a modular structure separating in classes the different concepts present in \RL like the environment, the observations and rewards, the actions, etc. This modularity allowed us to extend the different classes and ease their adaptation to our problem. The \textit{PyBrain} also provides flexibility to configure the different parameters of the \RL algorithm.

\subsection{Simple Example}
\label{sec:example}

This section uses a 2D convolution kernel as an example to show the resulting \textit{state-action} table obtained after learning from a simple transformation sequence with five states and two transformation rules. The first rule ($R_0$) considered in the example transforms a non-1D array into a 1D array. The second rule ($R_1$) performs a collapse of two nested for loops producing a single loop. The initial, intermediate and final codes obtained from the application of these rules is described below. Listing~\ref{lst:convInitial} through Listing~\ref{lst:convStep4} highlights changes in code using \red{this style} for indicating the portion of the code that will be changed after rule application. Highlight code using {\color{mauve}this style} indicates the resulting code after applying some transformation rule.

Listing~\ref{lst:convInitial} shows the initial code and the associated vector of features, as a code comment %\stcommin{as a comment...}
, according to the description in Section~\ref{sec:abstraction}.

%\begin{lstlisting}[style=cstyle,caption=Initial Code,label=lst:convInitial]
\begin{lstlisting}[style=cstyle,caption=Initial Code,label=lst:convInitial]
// ABSTR: [3, 0, 0, 0, 0, 0, 1, 0, 0, 1, 1, 3, 2, 4, 0]

int dead_rows = K / 2;
int dead_cols = K / 2;

int normal_factor = K * K;

for (r = 0; r < N - K + 1; r++) {
  for (c = 0; c < N - K + 1; c++) {
    sum = 0;
    for (i = 0; i < K; i++) {
      for (j = 0; j < K; j++) {
        sum += input_image@\red{[r+i][c+j]}@ * kernel[i][j];
      }
    }
    output_image[r+dead_rows][c+dead_cols] = (sum / normal_factor);
  }
 }
\end{lstlisting}  
%\end{lstlisting}

Listing~\ref{lst:convStep1} shows the result of applying rule $R_0$ to the code in Listing~\ref{lst:convInitial}. It can be seen that the array \texttt{input\_image} is transformed into a one dimensional array and the vector of features associated to code changes accordingly.

\begin{lstlisting}[style=cstyle,caption=Transformation step 1,label=lst:convStep1]
// ABSTR: [3, 0, 0, 0, 0, 0, 1, 0, 0, 1, 1, @{\color{mauve}2}@, 2, 4, 0]

int dead_rows = K / 2;
int dead_cols = K / 2;

int normal_factor = K * K;

for (r = 0; r < N - K + 1; r++) {
  for (c = 0; c < N - K + 1; c++) {
    sum = 0;
    for (i = 0; i < K; i++) {
      for (j = 0; j < K; j++) {
        sum += input_image@{\color{mauve} [(r+i)*(N - K + 1) + (c+j)]}@ * @\red{kernel[i][j]}@;
      }
    }
    output_image[r+dead_rows][c+dead_cols] = (sum / normal_factor);
  }
 }
\end{lstlisting}

Listing~\ref{lst:convStep2} shows the result of applying again rule $R_0$ to the code in Listing~\ref{lst:convStep1}. It can be seen that now the array \texttt{kernel} is transformed into a one dimensional array and the vector of features associated to code changes accordingly.

%\newpage

\begin{lstlisting}[style=cstyle,caption=Transformation step 2,label=lst:convStep2]
// ABSTR: [3, 0, 0, 0, 0, 0, 1, 0, 0, 1, 1, @{\color{mauve}1}@, 2, 4, 0]

int dead_rows = K / 2;
int dead_cols = K / 2;

int normal_factor = K * K;

for (r = 0; r < N - K + 1; r++) {
  for (c = 0; c < N - K + 1; c++) {
    sum = 0;
    for (i = 0; i < K; i++) {
      for (j = 0; j < K; j++) {
        sum += input_image[(r+i)*(N - K + 1) + (c+j)] * @{\color{mauve} kernel[i*K+j]}@;
      }
    }
    output_image@\red{[r+dead\_rows][c+dead\_cols]}@ = (sum / normal_factor);
  }
 }
\end{lstlisting}  

Listing~\ref{lst:convStep3} shows the result of applying rule $R_0$ to the code in Listing~\ref{lst:convStep2}. It can be seen that the array \texttt{output\_image} is transformed into a one dimensional array and the vector of features associated to code changes accordingly.

\begin{lstlisting}[style=cstyle,caption=Transformation step 3,label=lst:convStep3]
// ABSTR: [3, 0, 0, 0, 0, 0, 1, 0, 0, 1, 1, @{\color{mauve}0}@, 2, 4, 0]

int dead_rows = K / 2;
int dead_cols = K / 2;

int normal_factor = K * K;

@\red{for(r = 0; r < N - K + 1; r++)}@ {
  @\red{for(c = 0; c < N - K + 1; c++)}@ {
    sum = 0;
    for (i = 0; i < K; i++) {
      for (j = 0; j < K; j++) {
        sum += input_image[(@\red{r}@+i)*(N - K + 1) + (@\red{c}@+j)] * kernel[i*K+j];
      }
    }
    output_image@{\color{mauve}[(\red{r}+dead\_rows)*(N-K+1) + (\red{c}+dead\_cols)]}@ = (sum / normal_factor);
  }
 }
\end{lstlisting}

Listing~\ref{lst:convStep4} shows the result of applying rule $R_1$ to the code in Listing~\ref{lst:convStep3}. It can be seen that the two outermost loops are collapsed into one for loop but keeping an iteration space with the same number of iterations. Now the code abstraction reflects the change since the number or loops has decreased by one.

\begin{lstlisting}[style=cstyle,caption=Transformation step 4,label=lst:convStep4]
// ABSTR: [@{\color{mauve}2}@, 0, 0, 0, 0, 0, 1, @{\color{mauve}1}@, 0, 1, 1, 0, 2, 4, 0]

int dead_rows = K / 2;
int dead_cols = K / 2;

int normal_factor = K * K;

@{\color{mauve} for (z = 0; z < (N - K + 1)*(N - K + 1); z++)}@ {
    sum = 0;
    for (i = 0; i < K; i++) {
      for (j = 0; j < K; j++) {
        sum += input_image[(@{\color{mauve} (z / (N - K + 1))}@+i)*(N - K + 1) + (@{\color{mauve} (z \% (N - K + 1))}@+j)] * kernel[i*K+j];
      }
    }
    output_image[(@{\color{mauve} (z / (N - K + 1))}@+dead_rows)*(N - K + 1) +
	       (@{\color{mauve} (z \% (N - K + 1))}@+dead_cols)] = (sum / normal_factor);
}
\end{lstlisting}

Table~\ref{tbl:learningTable} shows the resulting values of the \textit{state-action} table ($Q$) for the simple transformation sequence described before. In Table~\ref{tbl:learningTable} there are as many rows as states obtained from the evaluation of $SM(A(C_i))$ for each code $C_i$, where $C_0$ is the initial code and $C_4$ is the final code classified as ready code for FPGA. Table~\ref{tbl:learningTable} shows the learned sequence composed of four steps: three consecutive applications of rule $R_0$ and one application of rule $R_1$. The $Q$ values of this sequence are highlighted in \grey{blue}. It can be seen how $Q$ values decreases from the state $SM(A(C_3))$, with the highest reward, down to the initial state $SM(A(C_0))$. This decay behavior in the sequence where $Q$ values decrease as states get further from final states, is caused by the discount factor ($\gamma$) introduced in Equation~\ref{eq:RL}. It should be noted that $Q$ values for final states are not updated by the recursive expression in Equation~\ref{eq:RL}. Thus, the final state $SM(A(C_4))$ keeps the initial value of 1.

\begin{table}[!h]
\begin{tabular}{ c|c|c||c| } 
\cline{2-4}
                                                            & $AM(R_0)$                    & $AM(R_1)$ & $RS(C_i)$ \\ \hline
 \multicolumn{1}{ |c| }{$SM(A(C_0))$} &  \grey{17.03718317} &  16.21544456 & $R_0$ \\
 \multicolumn{1}{ |c| }{$SM(A(C_1))$} &  \grey{17.25327145} & 16.80486418 & $R_0$ \\
 \multicolumn{1}{ |c| }{$SM(A(C_2))$} &  \grey{17.51541052} & 16.7189079 & $R_0$ \\
 \multicolumn{1}{ |c| }{$SM(A(C_3))$} &  16.72942327 & \grey{17.78007298} & $R_1$ \\
 \multicolumn{1}{ |c| }{$SM(A(C_4))$} &  1.                  &    1.                          & - \\  

 \hline
\end{tabular}
\caption{Values learned for $Q$ table}
\label{tbl:learningTable}
\end{table}

The example used in this section shows the transformation of a piece of C code. However, the fact that the \ML methods used work on program abstractions makes the approach generic and suitable for other imperative languages like FORTRAN, which is also widely used in scientific computing. The application of the approach to other languages would require changes to the tool described in Section~\ref{sec:abstraction} in order to account for some specific syntactical patterns of a particular programming language. Nevertheless, most of the abstraction features identified and described in Section~\ref{sec:abstraction} are also applicable to other imperative languages since they account for common aspects like: control flow, data layout, data dependencies, etc.

\section{Results}
\label{sec:Results}

This section shows some preliminary results obtained for the \ML-based transformation strategies. We have performed some measurements to verify our claims in Section~\ref{sec:RL} regarding the non-monotonic behavior of non-functional properties for codes obtained from a transformation sequence and how transformed codes with non-optimal values can produce better performant final codes. In order to show this behavior we have identified four different transformation sequences for a use case application performing image compression using the discrete cosine transform.

\begin{figure}[!h]
\center
\includegraphics[width=0.4\textwidth]{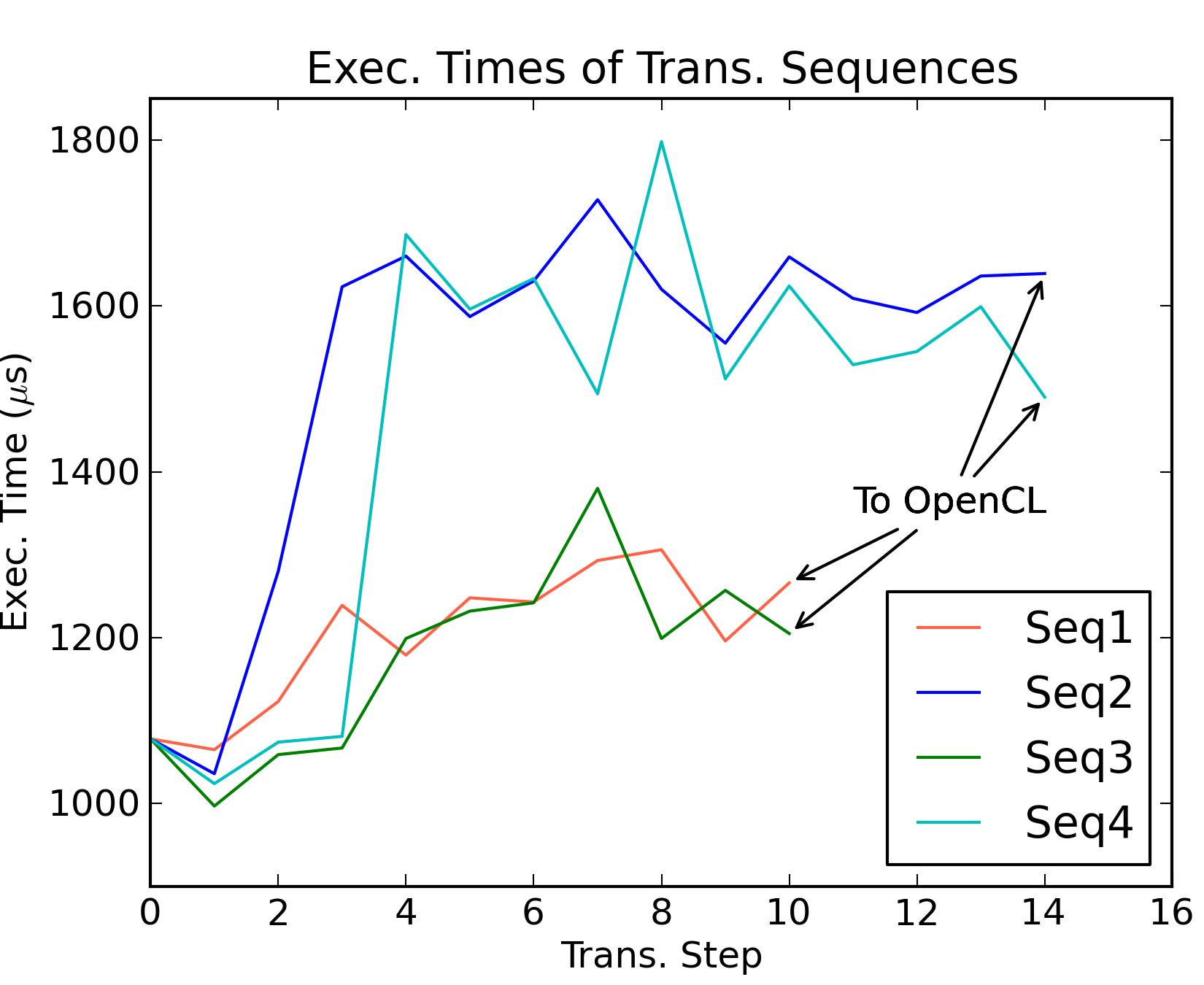}
\caption{Execution times for transformation sequences}
\label{fig:seqExecTime}
\end{figure}

Each of the identified sequences produces C code which is adequately shaped to be mechanically translated to OpenCL and executed on a GPU. First, we have measured the average execution time of 30 runs for each intermediate code of each sequence. These results are reported in Figure~\ref{fig:seqExecTime} which shows the evolution of the execution times. It can be seen that execution times of sequences do not follow a monotonic behavior and have constant changes. As expected, we also obtain different sequential codes \textit{ready} to be translated to OpenCL and with different execution times.

\begin{figure}[!h]
\center
\includegraphics[width=0.4\textwidth]{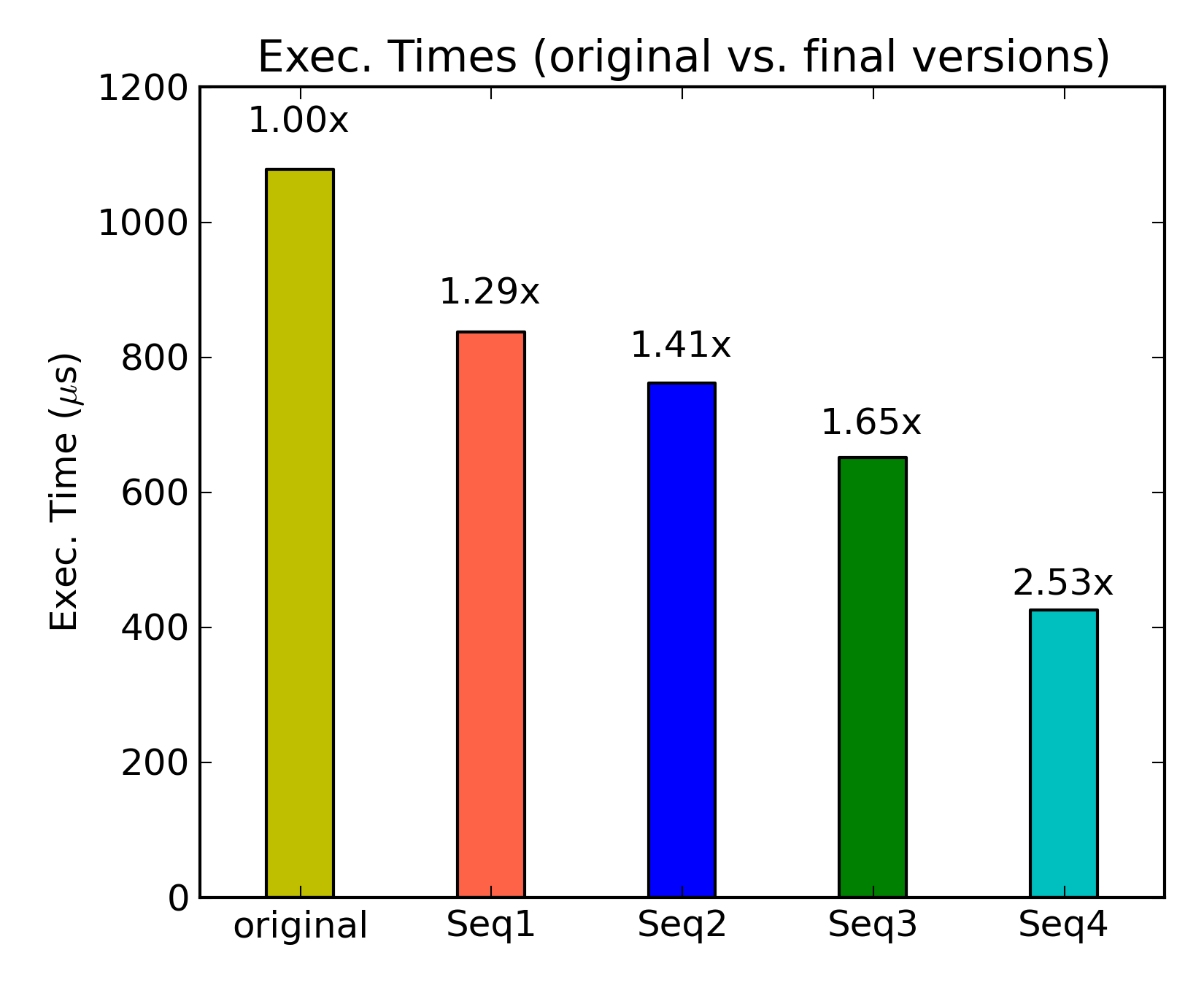}
\caption{Execution times for OpenCL versions}
\label{fig:finalExecTime}
\end{figure}

As a second step, we have measured the performance of the OpenCL version generated from each sequence. Figure~\ref{fig:finalExecTime} shows the execution times of the original sequential program and OpenCL versions. Figure~\ref{fig:finalExecTime} also reports, on top of each bar, the speed up of each OpenCL version with respect to the sequential program, showing that the version obtained from sequence 4 is the fastest one accelerating the computation by a factor of 2.53x. Now, looking at Figures~\ref{fig:seqExecTime} and~\ref{fig:finalExecTime} together it can be seen the absence of correlation between the non-functional property measured (i.e. execution time) of each final sequential code and the performance of the corresponding OpenCL version. In fact, sequence 4 with the second highest execution time produces the fastest GPU implementation. Based on these results it can be concluded that an effective method must be used to discover and learn the uncorrelated relation between final sequential codes and parallel versions for a range of target platforms. We have decided to base our approach on \RL, since it is driven by final performance measurements rather than on intermediate values which can lead to suboptimal results.

We have also performed a preliminary evaluation of \RL as a technique to guide a rule-based program transformation tool. In order to do the aforementioned evaluation we have selected four use case applications and identified different transformation sequences leading to codes that can be mechanically translated to OpenCL and executed on a GPU. These four applications and its corresponding transformation sequences have been used as training set. One of the applications of the train set is the image compression program, mentioned before, denoted as \textit{compress}. Other program of the train set, denoted as \textit{rgbFilter}, is an image filter that splits the different color channels of a given RGB input image in different images. Two more image processing applications complete the train set: one, denoted as \textit{edgeDetect}, detects edges in an image using a Sobel filter and another one, denoted as \textit{threshold}, performs image segmentation over an input image given a threshold value.  Once the training set is defined, the  transformation engine based on \RL requires to tune the two parameters that appear in Equation~\ref{eq:RL}, i.e. the learning rate ($\alpha$) and discount factor ($\gamma$). For this purpose, an empirical study was performed in which parameter values leading to transformation sequences providing the fastest OpenCL versions were selected.
% by checking whether the transformation sequences learnt for each application were the ones providing the best results, i.e. the fastest OpenCL versions. 
According to this criteria, a value of 0.5 was used for $\alpha$ and 0.6 for $\gamma$. Also, reward values have been chosen in order to give a higher reinforcement to those sequences leading to better performant final codes. In our problem, reward values used for the best sequences of each use case application are substantially higher, with a ratio of 1:100, with respect to the rest of transformation sequences.

After training,
%% the Reinforcement Learning-based engine described in
%% Section~\ref{sec:Learning} was trained using the program set with
%% four applications described above,
three different use case applications were used as predict set. The applications in the predict set were mechanically transformed according to the previously learned sequences and finally translated by hand into OpenCL.  Independently, OpenCL versions of the initial C code were written.  The evaluation of the the \RL approach was made
%% by hand coding OpenCL versions of each train and predict
%% application. In this way, we can perform our evaluation
by comparing the performance of the hand-coded versions with that of the versions mechanically generated from the transformed codes produced by the sequences learnt by \RL.

\begin{figure}[!h]
\center
\includegraphics[width=0.4\textwidth]{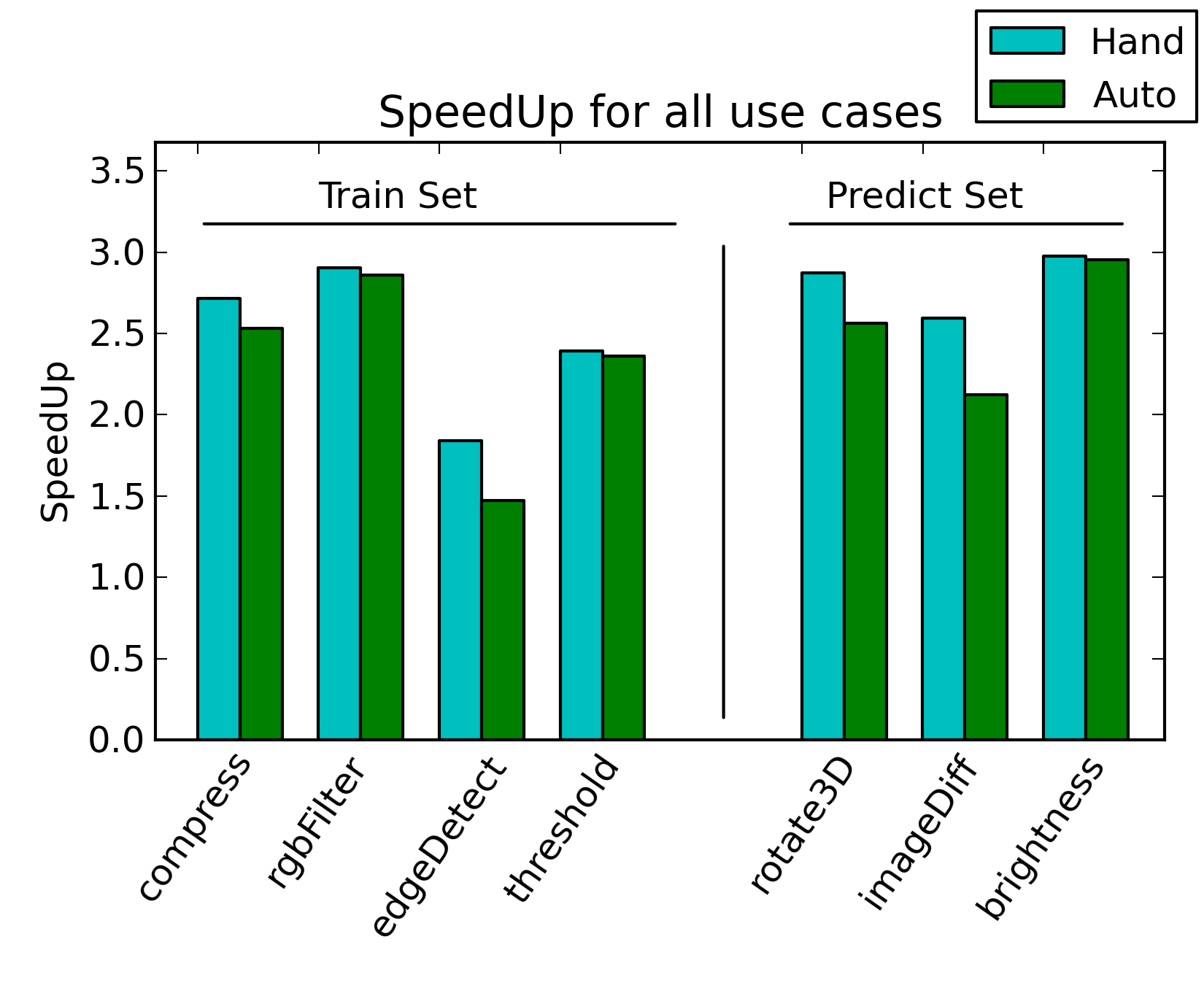}
\caption{Speed up for train and predict sets}
\label{fig:allSpeedUp}
\end{figure}

Figure~\ref{fig:allSpeedUp} shows the results obtained for both the train and predict sets in terms of speed up factor of the hand and mechanically generated OpenCL versions over the original sequential program. By comparing the results for both the train and the predict set we can assess the response of the transformation system after the learning phase. Looking at results in Figure~\ref{fig:allSpeedUp} we can see that the transformation sequences leads to parallel versions that provide comparable acceleration factors with respect to hand coded versions. Although these preliminary evaluation is based on a small sample, it shows that our approach seems promising to tackle the problem of rule-based program transformation systems.

The results discussed in this section show a preliminary evaluation for GPU platforms. However, the same approach can be followed in order to target different platforms which are present in Heterogeneous systems. In this way, a separate \textit{state-action} table can be used for learning adequate transformation sequences for each target platform and rewards would be assigned based on the performance of the final code generated for each platform.

%\appendix
%\section{Appendix Title}
%
%This is the text of the appendix, if you need one.

\section{Conclusions and Future Work}
\label{sec:conclusions}

In this paper we have proposed a \ML-based approach to learn heuristics for guiding the code transformation process of a rule-based program
transformation system \cite{tamarit15:padl-haskell_transformation,tamarit16:code_trans}. This type of systems pose a number of problems such as the search-space exploration problem arising from the application of transformation rules in arbitrary orders or the definition of a stop criteria of the transformation system. For the latter we propose the use of classification trees and for the former we propose a novel approach based on \RL.  We have also performed a preliminary evaluation of the approach, which provided promising results that demonstrate the suitability of the approach for this type of transformation systems.

%\gvcommin{Mention to use more non-functional properties, like energy, to define rewards in RL}

As a future work we plan to continue expanding the set of use case applications considered for training the different \ML techniques used. As a consequence, we expect to enrich the code features identified for obtaining program abstractions in order to capture new code patterns. We also expect that the increase in the training set will result in better prediction outcomes, but it will also increase the complexity of efficiently using the learning techniques. In the case of \RL having more learning cases results in a bigger state space.
%\mclcommin{Not sure I understand the previous sentence?}
 For that reason we plan to use common methods to reduce the number of states like clustering or principal component analysis techniques. We also plan to explore another features of \RL like the use of different learning rates for different states or transformation sequences in order to learn and converge faster towards transformed codes providing better performant final versions. 

We also plan to define \textit{multi-objective} \RL rewards based on different non-functional properties like energy consumption, resource usage, etc. and even a combination of multiple of these properties. This future line would permit the definition of transformation strategies that try to optimize different aspects and generate, for example, a final code among the fastest ones that consumes the least amount of energy. Also, the introduction of weights into the \textit{multi-objective} rewards would offer programmers the flexibility to select which non-functional property or set of properties they want to focus on for generating the final code.

\acks

Work partially funded by EU FP7-ICT-2013.3.4 project 610686 POLCA, Comunidad de Madrid project S2013/ICE-2731 N-Greens Software, and MINECO Projects TIN2012-39391-C04-03 / TIN2012-39391-C04-04 (StrongSoft), TIN2013-44742-C4-1-R (CAVI-ROSE), and TIN2015-67522-C3-1-R (TRACES).

% We recommend abbrvnat bibliography style.

\bibliographystyle{abbrvnat}
% The bibliography should be embedded for final submission.
\bibliography{../../BiBTeX/hpc_transformations,../../BiBTeX/polca_refs,../../BiBTeX/machine_learning}

%\begin{thebibliography}{}
%\softraggedright
%
%\bibitem[Smith et~al.(2009)Smith, Jones]{smith02}
%P. Q. Smith, and X. Y. Jones. ...reference text...
%
%\end{thebibliography}

\end{document}